\documentclass{aa}

\usepackage{graphicx}
\usepackage{geometry}
\usepackage{txfonts}
\usepackage{amsmath, amsfonts, amssymb}
\usepackage{xcolor}
\usepackage{caption, subcaption}
\usepackage{ulem}
\usepackage{url}

\usepackage{hyperref}
\usepackage{todonotes}
\usepackage{sansmath}
\usepackage{csvsimple}
\usepackage{multirow}

\bibpunct{(}{)}{;}{a}{}{,}

\DeclareRobustCommand{\VAN}[3]{#2}
\let\VANthebibliography\thebibliography
\def\thebibliography{\DeclareRobustCommand{\VAN}[3]{##3}\VANthebibliography}

\def\cm2{\rm \ cm$^{-2}$}

\def\src{EP\,J1154$-$5018}

\begin{document}

%Bibliography and bibfile
\def\aj{AJ}%
          % Astronomical Journal
\def\actaa{Acta Astron.}%
          % Acta Astronomica
\def\araa{ARA\&A}%
          % Annual Review of Astron and Astrophys
\def\apj{ApJ}%
          % Astrophysical Journal
\def\apjl{ApJ}%
          % Astrophysical Journal, Letters
\def\apjs{ApJS}%
          % Astrophysical Journal, Supplement
\def\ao{Appl.~Opt.}%
          % Applied Optics
\def\apss{Ap\&SS}%
          % Astrophysics and Space Science
\def\aap{A\&A}%
          % Astronomy and Astrophysics
\def\aapr{A\&A~Rev.}%
          % Astronomy and Astrophysics Reviews
\def\aaps{A\&AS}%
          % Astronomy and Astrophysics, Supplement
\def\azh{AZh}%
          % Astronomicheskii Zhurnal
\def\baas{BAAS}%
          % Bulletin of the AAS
\def\bac{Bull. astr. Inst. Czechosl.}%
          % Bulletin of the Astronomical Institutes of Czechoslovakia
\def\caa{Chinese Astron. Astrophys.}%
          % Chinese Astronomy and Astrophysics
\def\cjaa{Chinese J. Astron. Astrophys.}%
          % Chinese Journal of Astronomy and Astrophysics
\def\icarus{Icarus}%
          % Icarus
\def\jcap{J. Cosmology Astropart. Phys.}%
          % Journal of Cosmology and Astroparticle Physics
\def\jrasc{JRASC}%
          % Journal of the RAS of Canada
\def\mnras{MNRAS}%
          % Monthly Notices of the RAS
\def\memras{MmRAS}%
          % Memoirs of the RAS
\def\na{New A}%
          % New Astronomy
\def\nar{New A Rev.}%
          % New Astronomy Review
\def\pasa{PASA}%
          % Publications of the Astron. Soc. of Australia
\def\pra{Phys.~Rev.~A}%
          % Physical Review A: General Physics
\def\prb{Phys.~Rev.~B}%
          % Physical Review B: Solid State
\def\prc{Phys.~Rev.~C}%
          % Physical Review C
\def\prd{Phys.~Rev.~D}%
          % Physical Review D
\def\pre{Phys.~Rev.~E}%
          % Physical Review E
\def\prl{Phys.~Rev.~Lett.}%
          % Physical Review Letters
\def\pasp{PASP}%
          % Publications of the ASP
\def\pasj{PASJ}%
          % Publications of the ASJ
\def\qjras{QJRAS}%
          % Quarterly Journal of the RAS
\def\rmxaa{Rev. Mexicana Astron. Astrofis.}%
          % Revista Mexicana de Astronomia y Astrofisica
\def\skytel{S\&T}%
          % Sky and Telescope
\def\solphys{Sol.~Phys.}%
          % Solar Physics
\def\sovast{Soviet~Ast.}%
          % Soviet Astronomy
\def\ssr{Space~Sci.~Rev.}%
          % Space Science Reviews
\def\zap{ZAp}%
          % Zeitschrift fuer Astrophysik
\def\nat{Nature}%
          % Nature
\def\iaucirc{IAU~Circ.}%
          % IAU Cirulars
\def\aplett{Astrophys.~Lett.}%
          % Astrophysics Letters
\def\apspr{Astrophys.~Space~Phys.~Res.}%
          % Astrophysics Space Physics Research
\def\bain{Bull.~Astron.~Inst.~Netherlands}%
          % Bulletin Astronomical Institute of the Netherlands
\def\fcp{Fund.~Cosmic~Phys.}%
          % Fundamental Cosmic Physics
\def\gca{Geochim.~Cosmochim.~Acta}%
          % Geochimica Cosmochimica Acta
\def\grl{Geophys.~Res.~Lett.}%
          % Geophysics Research Letters
\def\jcp{J.~Chem.~Phys.}%
          % Journal of Chemical Physics
\def\jgr{J.~Geophys.~Res.}%
          % Journal of Geophysics Research
\def\jqsrt{J.~Quant.~Spec.~Radiat.~Transf.}%
          % Journal of Quantitiative Spectroscopy and Radiative Trasfer
\def\memsai{Mem.~Soc.~Astron.~Italiana}%
          % Mem. Societa Astronomica Italiana
\def\nphysa{Nucl.~Phys.~A}%
          % Nuclear Physics A
\def\physrep{Phys.~Rep.}%
          % Physics Reports
\def\physscr{Phys.~Scr}%
          % Physica Scripta
\def\planss{Planet.~Space~Sci.}%
          % Planetary Space Science
\def\procspie{Proc.~SPIE}%
          % Proceedings of the SPIE
\let\astap=\aap
\let\apjlett=\apjl
\let\apjsupp=\apjs
\let\applopt=\ao

\title{Einstein Probe discovery of the short period intermediate polar EP\,J115415.8$-$501810}
\author{Y.~Xiao\inst{1}\thanks{\url{xiaoyunxiang@ihep.ac.cn, gemy@ihep.ac.cn, rea@ice.csic.es}}
          \and M.~Ge\inst{1}
          \and N.~Rea\inst{2,3}
          \and F.~Lu\inst{1}
          \and H.~Feng\inst{1}
          \and L.~Tao\inst{1}
          \and D.~de~Martino\inst{4}
          \and F. Coti Zelati\inst{2,3} 
          \and \\ A. Marino\inst{2,3} 
          \and E. Kuulkers\inst{5}
          \and W. Yuan\inst{6,7}
          \and C. Jin\inst{6,7,19}
          \and H. Sun\inst{6}
          \and J. Wu\inst{8}
          \and N. Hurley-Walker\inst{9}
          \and S. J. McSweeney\inst{9} 
          \and \\ D. A. H. Buckley\inst{10,11,12}
          \and B. Zhang\inst{13,14} 
          \and S. Zhang\inst{1,7} 
          \and S. Scaringi\inst{15,4}
          \and K. Mori\inst{16}
          \and Z.~Yu\inst{17}
          \and X.~Hou\inst{18}
          \and Y. Xu\inst{1}
          }

   \institute{Key Laboratory of Particle Astrophysics, Institute of High Energy Physics, Chinese Academy of Sciences, Beijing 100049, China
   \and
   Institute of Space Sciences (ICE-CSIC), Campus UAB, C/ de Can Magrans s/n, Cerdanyola del Vallès (Barcelona) 08193, Spain
   \and
   Institut d'Estudis Espacials de Catalunya (IEEC), Esteve Terradas 1, RDIT Building, Of. 212 Mediterranean Technology Park (PMT), 08860, Castelldefels, Spain
   \and INAF -- Osservatorio Astronomico di Capodimonte, Salita Moiarello 16 80131, Napoli, Italy
   \and ESA/ESTEC, Noordwijk, 2201 AZ, The Netherlands
   \and National Astronomical Observatories, Chinese Academy of Sciences, Beijing, 100101, People’s Republic of China
   \and School of Astronomy and Space Sciences, University of Chinese Academy of Sciences, Beijing, 100049, People’s Republic of China
   \and Department of Astronomy, Xiamen University, Xiamen, Fujian 361005, China
   \and International Centre for Radio Astronomy Research, Curtin University, Kent Street, Bentley WA, 6102, Australia
   \and South African Astronomical Observatory, PO Box 9, Observatory 7935, South Africa
   \and Department of Astronomy, University of Cape Town, Private Bag X3, Rondebosch 7701, South Africa
    \and Department of Physics, University of the Free State, PO Box 339, Bloemfontein 9300, South Africa
    \and The Nevada Center for Astrophysics, University of Nevada, Las Vegas, Las Vega , 89154, NV, USA
    \and Department of Physics and Astronomy, University of Nevada, Las Vegas, Las Vega , 89154, NV, USA
   \and Centre for Extragalactic Astronomy, Department of Physics, Durham University, South Road, Durham, DH1 3LE
   \and Columbia Astrophysics Laboratory, Columbia University, New York, NY 10027, USA
   \and College of Physics and Electronic Engineering, Qilu Normal University, 250200, Jinan, People’s Republic of China
   \and Yunnan Observatories, Chinese Academy of Sciences, Kunming 650216, People’s Republic of China
   \and Institute for Frontier in Astronomy and Astrophysics, Beijing Normal University, Beijing, 102206, People’s Republic of China}

   \date{Received XXX 00, 2025; accepted YYY 00, 2025}

\abstract{

The X-ray transient source EP240309a/EP\,J115415.8$-$501810 was first detected by the Wide-Field X-ray Telescope (WXT) on board Einstein Probe (EP) during the commissioning phase. Subsequent optical observations confirmed it as a Cataclysmic Variable of the intermediate polar type with a 238.2\,s spinning white dwarf in a $\sim$3.76\,hr orbit. We report on the source discovery and follow-up studies made with the Follow-up X-ray Telescope (FXT) of EP. A periodic variation of 231\,s is detected in the 0.3$-$2\,keV band, while no obvious pulsation appears in the 2$-$10\,keV band. The spectral analysis shows that the X-ray emission could be described by an absorbed bremsstrahlung model with $kT$\textgreater\,11\,keV. The partial covering absorption, with an hydrogen column density $N_H$ = 2.0$\times 10^{22}\,\rm cm^{-2}$ and covering fraction around 0.9, is much larger than the interstellar absorption along the line of sight. According to the distance $d = 309.5$\,pc obtained from Gaia parallax, we estimate that the luminosity of this source in the 0.3$-$10\,keV range is $\sim 2\times10^{32}$\,erg\,s$^{-1}$. In addition, phase-resolved spectral analysis reveals that the detected periodic variation is mainly caused by the change in the absorption column density. In this scenario the spin modulation arises due to absorption from the pre-shock accretion flow of the X-ray emitting pole, while the optical radiation is modulated at the orbital side band ($\omega_{\rm spin} - \Omega_{\rm orbit}$) due to reprocessing in regions within the binary system. Due to its unusual transient behaviour for an intermediate polar, we have also searched for radio signals similar to those observed in the new class of long period transients. We derived upper limits with ASKAP (200--300\,$\mu$Jy\,beam$^{-1}$ between 800--1500 MHz) and MWA (40--90\,mJy\,beam$^{-1}$ between 80--300 MHz).

}


\maketitle

\section{Introduction} \label{sec:intro}

White dwarfs (WD), the common endpoints of the evolutions of solar-mass stars, are often observed in binary systems known as Cataclysmic Variables (CVs), where they accrete material from a low-mass companion. Intermediate polars (IPs) are one kind of magnetic CVs in which the WD accretes mass from a late-type main-sequence star and spins faster than the orbital period \citep{1994PASP..106..209P, 2017PASP..129f2001M}. Depending on the WD magnetic field strength, the accreted matter might form accretion disks truncated at the magnetospheric radius, from which the material is funneled onto the polar caps along the magnetic field lines, releasing gravitational energy. A stand-off shock is formed at the pole(s) below which matter cools via optically thin thermal ($\sim$ 10-40\, keV) X-ray radiation giving rise to pulsed emission at the rotational period of the WD. Modulations of the X-ray emission at the orbital period are rarely observed in relatively high inclination systems \citep{2005A&A...439..213P,2018MNRAS.478.1185B}

\begin{table*}
\caption{ 
\label{tab:log}
List of the 2024 observations of \src\ presented in this work.}
\centering
\begin{tabular}{lccc}
\hline\hline
Telescope/Instrument	& Obs ID 		&Start -- End time	        		   & Exposure  \\
					   &			& Mmm DD hh:mm:ss		& (ks)    \\
\hline
EP/WXT           & 13600005097      & Mar 6  01:17:53 -- Mar 6  12:04:09         & 18.34
    \\
                 & 08500000016      & Mar 9  01:33:26 -- Mar 9  20:28:40         & 33.37
    \\
                 & 13600005117      & Mar 13 08:15:56 -- Mar 14 08:18:22         & 39.03
    \\
                 & 13600005118      & Mar 14 08:18:22 -- Mar 14 22:03:03         & 25.01
    \\
                 & 13600005120      & Mar 15 18:28:25 -- Mar 16 04:54:28         & 15.42
    \\
                 & 13600005121      & Mar 16 05:18:46 -- Mar 16 14:48:23         & 19.93
    \\
                 & 13600005122      & Mar 16 14:58:58 -- Mar 17 00:10:42         & 16.27
    \\
                 & 13600005155      & Apr 8  12:14:12 -- Apr 9  11:44:06         & 39.29
    \\
EP/FXT           & 08500000024      & Mar 16 02:23:24 -- Mar 16  03:14:58         & 3.09
    \\
\hline
\hline
\end{tabular}

\end{table*}

The X-ray  spectra of IPs are generally complex, consisting of heavily absorbed multi-temperature plasma and often also an optically thick component arising from the heated WD polar cap \citep{2017PASP..129f2001M, 2020AdSpR..66.1209D}. As of 2021, 71 IPs have been identified, and more than 100 additional candidates are known \footnote{\url{https://asd.gsfc.nasa.gov/Koji.Mukai/iphome/iphome.html}}.

On 2024 March 9, a new transient triggered the Wide-field X-ray Telescope (WXT) aboard the Einstein Probe (EP) during its commissioning phase and was initially called EP240309a \citep{2024ATel16546....1L}. Subsequent observations with EP's Follow-up X-ray Telescope (FXT), identified this new transient as a candidate CV in high state \citep{2024ATel16546....1L}, naming it EP\,J115415.8$-$501810 (\src\, hereafter). Follow-up studies, including observations from TESS and telescopes at the South African Astronomical Observatory, have revealed it to be an IP with a spin period of 3.97\,min and an orbital period of 3.76\,hr \citep{2024ATel16554....1B,2024MNRAS.532L..21P}. \cite{2024ATel16572....1C} also performed a radio observation at 1.28\,GHz and no radio counterpart was found within the error circle of the EP X-ray position. In this work, we report on the EP discovery of \src\, via detailed X-ray timing and spectral analyses. In Section \ref{sec:ep}, we present the data processing and data analysis results. In Section \ref{sec:Radio}, we present the radio upper limits. We discuss our results and draw a brief conclusion in Section \ref{sec:conclusion}.

\begin{figure}
\centering
\includegraphics[width=0.48\textwidth]{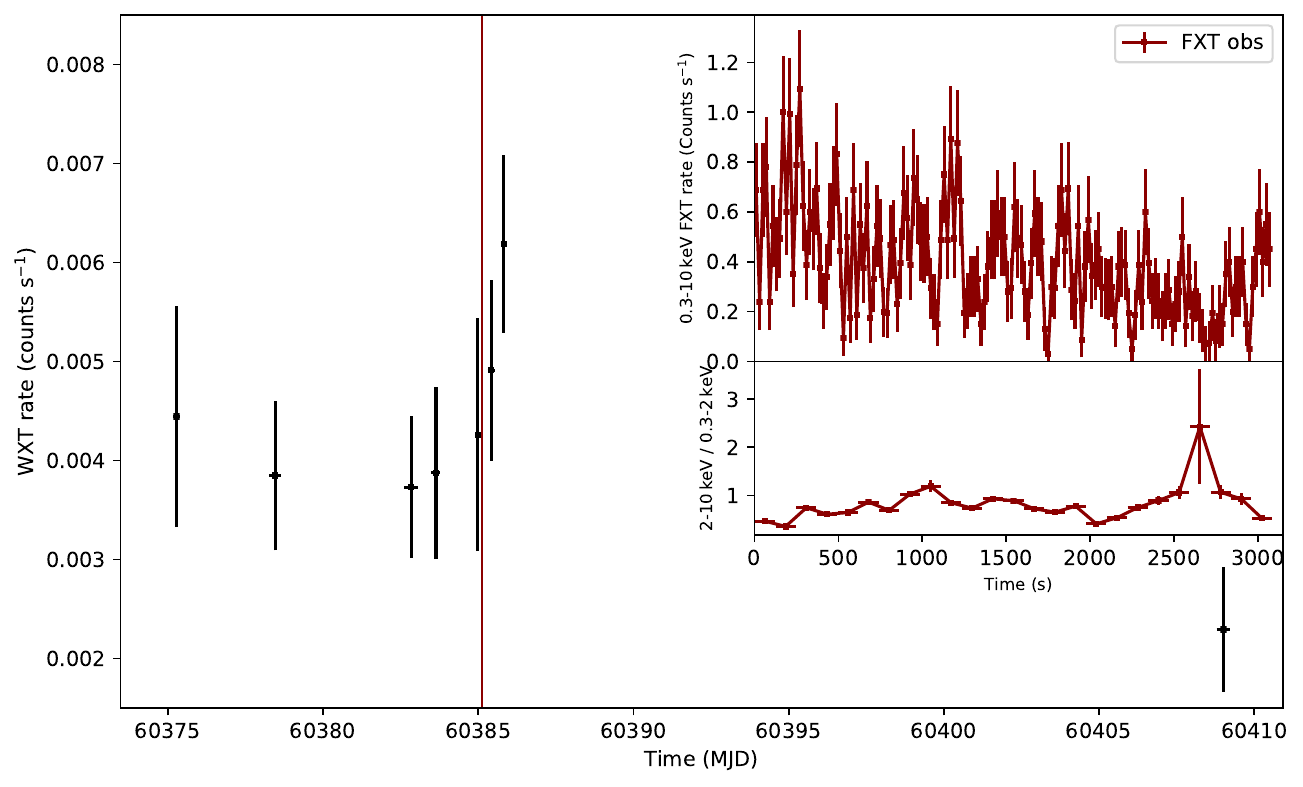}
\caption{Light curves of \src. The main panel shows the average count rates (black dots) of each observation by EP/WXT. The red vertical line indicates the start time of the EP/FXT observation. The inset displays the light curve (with 20\,s bin size) and hardness ratio (with 120\,s bin size) of EP/FXT observation. The details of the observations can be found in Table \ref{tab:log}. 
}
\label{fig:lc}
\end{figure}

\section{EP data analysis and results}
\label{sec:ep}

The main scientific objective of the EP mission is to monitor and survey the sky in the soft X-ray band. EP was launched on January 9, 2024 and soon entered the commissioning and calibration phase, and then formally started its scientific mission in July 2024. The scientific payloads of EP include WXT and FXT. WXT has lobster-eye micro-pore X-ray focusing optics, giving it an instantaneous larger field-of-view (3600\,sq.\,deg.) compared to previous wide-field X-ray monitors. It operates in the 0.5$-$4\,keV band and also has very high sensitivity ($(2-3) \times 10^{-11}\,\rm erg\,\rm cm^{-2}\,\rm s^{-1}$ for 1\,ks exposure). FXT consists of two co-aligned units, FXT-A and FXT-B. Each unit adopts a Wolter-I nested telescope and pn-CCDs as the focal plane detectors. FXT has a field-of-view of $1^\circ \times 1^\circ$ and an angular resolution of 30$^{\prime\prime}$. The telescope has a detection energy range of 0.3$-$10\,keV, with an effective area of 600\,cm$^2$ (2 units) at 1.25\,keV and an energy resolution of 120\,eV (FWHM) \citep{2022hxga.book...86Y}.

\src\ was detected at the position of R.A.$=178.57^{\circ}$, Dec$=-50.29^{\circ}$ (J2000, with an uncertainty of 2.1 arcmin) \citep{2024ATel16546....1L}. There were 8 WXT observations covering the field of this source (see Table \ref{tab:log}). The WXT data were reduced using the \texttt{wxtpipeline} tool from the WXT Data Analysis Software (\texttt{WXTDAS}). The calibration processes of the raw data include coordinate transformation, flagging hot and bad pixels, calculating the pulse-invariant values, screening events and extracting a sky image. We used the tools in \texttt{WXTDAS} to detect point-like sources and extract light curves and spectra as well as the corresponding backgrounds, the ancillary response files and the response matrices. The default region for source extraction is a circle with a radius of $9.1^\prime$, and the background region is an annulus with the inner and outer radii of $18.2^\prime$ and $36.4^\prime$, respectively.

\begin{figure}
\centering
\includegraphics[width=0.5\textwidth]{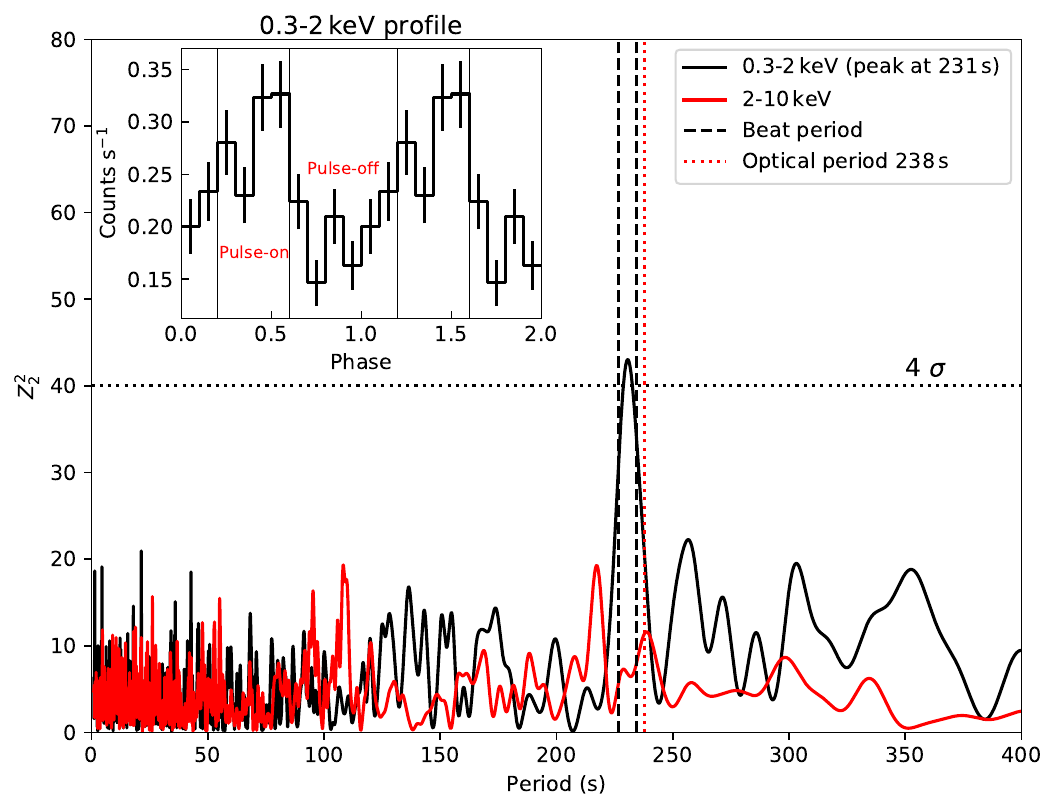}
\caption{Period search results and pulse profile of EP/FXT data. The main panel shows the period searching results using $Z^{2}_{2}$ test in energy bands of 0.3$-$2\,keV (black solid line) and 2$-$6\,keV (red solid line). The error on the best period value is estimated to be 2\,s (calculated by the bootstrap method). The dotted horizontal represents the 4\,$\sigma$ significance level obtained by the bootstrap method. We also mark the negative and positive beat period with black dashed line and mark the optical spin period \citep{2024MNRAS.532L..21P} with red dot line. The inset shows the $0.3-2$\,keV pulse profile folded on the period of 231\,s. }

\label{fig:search}
\end{figure}

The Follow-up X-ray Telescope (FXT) conducted a 3093\,s observation with the full-frame (FF) mode on Mar 16, 2024 (ObsID 08500000024). This mode provides full imaging capability with a time resolution of 50\,ms, which could be used to search for periodicities in the sub-second regime. More accurate position of \src\ has been obtained by FXT, R.A.$=178.566^{\circ}$ and Dec$=-50.303^{\circ}$ (J2000, with an uncertainty of about 10 arcsec) \citep{2024ATel16546....1L}. For this observation, we first used the \texttt{fxtchain} tool in the Follow-up X-ray Telescope Data Analysis Software (\texttt{FXTDAS}, version 1.10)\footnote{\url{http://epfxt.ihep.ac.cn/analysis}} package to reduce the FXT data. This task chain includes particle identification, calculating the pulse invariant values, flagging bad pixels and hot pixels, as well as screening of good time intervals. The source region was defined by a circle with a radius of 0.9$^\prime$, while the background region was an annulus with inner and outer radii of 0.9$^\prime$ and 2.9$^\prime$, respectively. Then, the spectra and light curves were extracted with the given source and background regions, along with the corresponding ancillary files and redistribution matrices. In addition, we used the \texttt{fxtbary} tool to correct the photon arrival times to the Solar system barycenter using the JPL DE405 ephemeris.

The 0.3$-$10\,keV light curve with a bin size of 20\,s is extracted based on the combined events (FXT-A and FXT-B). At the same time, the hardness ratio ($2-6$\,keV / $0.3-2$\,keV) in 120\,s bins is calculated in order to search for possible spectral variations. The 0.5-4\,keV light curve observed by WXT shows an increase of flux near the time of the FXT observation, as shown in Figure \ref{fig:lc}. During the FXT observation, the 0.3-10\,keV light curve illustrates a decreasing trend with relatively strong variations in short time scales but subsequent WXT observations show an increase in X-ray intensity (see Figure \ref{fig:lc}). At the same time, the hardness ratio shows a tendency to harden during the low-count-rate episodes. The FXT light curve can be fitted with a sine function at the known orbital period, indicating that the observed variations may be due to orbital motion. However, it cannot be verified due to limitations of the single exposure. On the other hand, the long-term WXT light curve shows an increase in intensity that suggests that the source is in a rise to a high state.

\begin{figure*}
\centering
\includegraphics[width=0.45\textwidth]{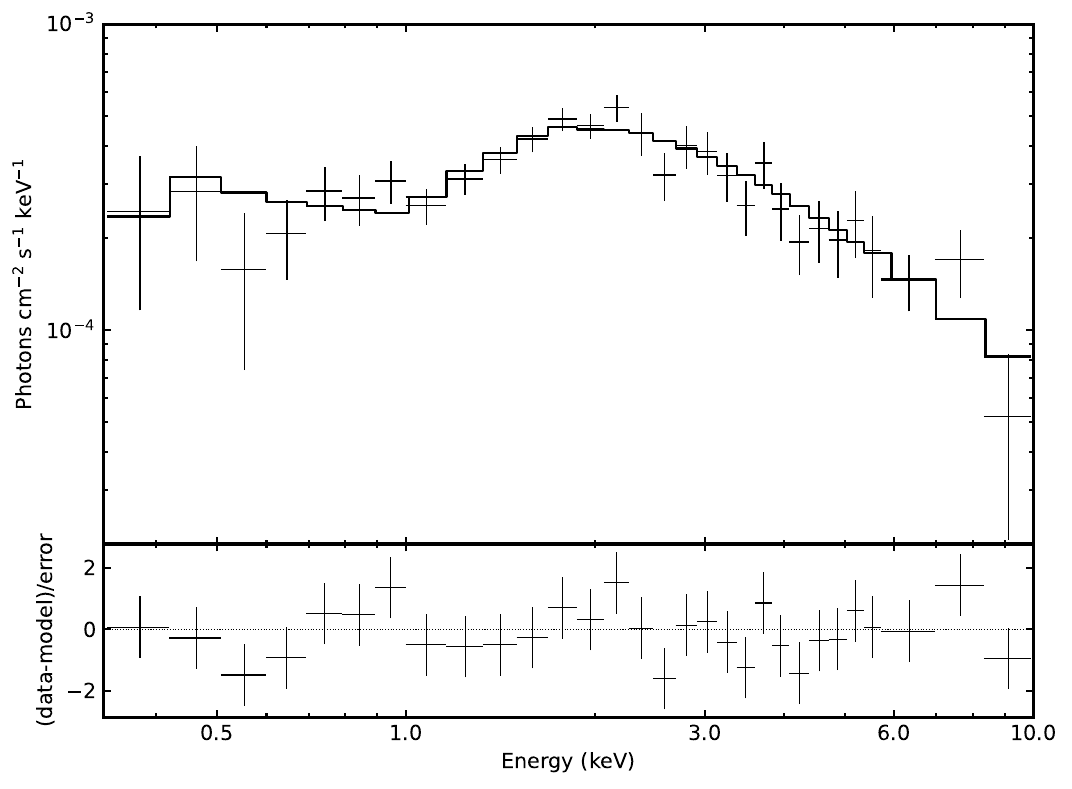}
\includegraphics[width=0.45\textwidth]{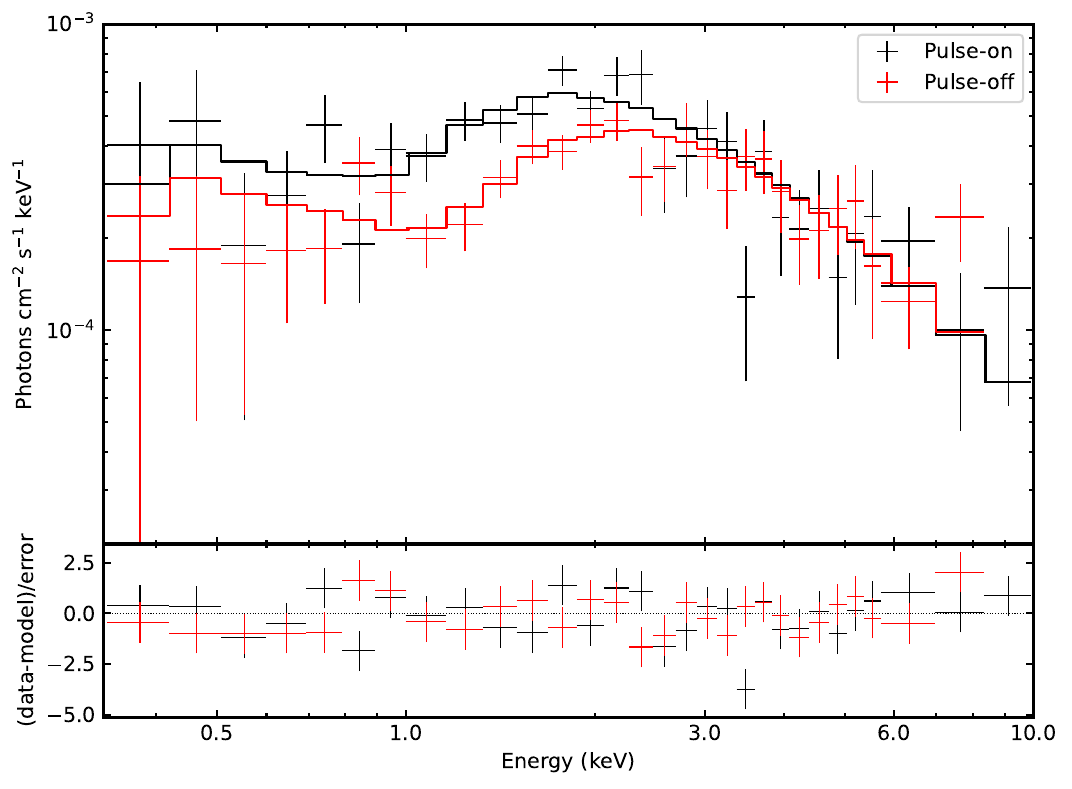}
\caption{Phase-average and phase-resolved spectral fitting of EP/FXT data. The left panel presents the  phase-average spectrum fitted with the model: $tbabs \times tbpcf \times bremss$ and the residuals of the fit. The right panel shows the phase-resolved spectra while "pulse-on" represents phase 0.2$-$0.6 and "pulse-off" represents phase 0.6$-$1.2 (see the inset in Figure \ref{fig:search}). The best-fitting parameters are listed in Table \ref{tab:par}.}
    \label{fig:spec}
\end{figure*}

For the FXT observation, we performed a periodic signal search in two energy bands, 0.3--2\,keV and 2--10\,keV, using the $Z_{2}^{2}$ test \citep{1983A&A...128..245B}, with the results shown in Figure \ref{fig:search}. We assume that the pulse profile is a superposition of two sine waves ($n=2$, taking a larger value of $n$ leads to similar result). We calculate the phase for each photon arrival time and obtain the value of $Z_{2}^{2}$. The $Z_{2}^{2}$ value peaks at the period of 231\,s for the $0.3-2$\,keV data, while no significant periodicity is detected based on the $2-10$\,keV data, which means that pulse signal was only significantly detected in the soft energy range. The result is consistent with that obtained by Pearson $\chi^2$ test, but with higher significance. Assuming that the potential variability is instrumental white noise, we use the period distribution searched from 10,000 samples based on the observed distribution, and estimate the 68\% confidence error to be 2\,s. Using the $Z_{2}^{2}$ distribution of 10,000 samples from a uniform distribution, the significance of this periodicity is found to be $\sim4\,\sigma$.  

Analysis of TESS data \citep{2024MNRAS.532L..21P} reported an orbital frequency $\Omega$ (period = 3.7614(4)\,hr). Assuming the spin frequency is $\omega_x$ (period = 231 $\pm$ 2\,s) as detected by FXT, then the periods corresponding to the positive ($\omega_x + \Omega$) and negative beat ($\omega_x - \Omega$) frequencies are 227\,s and 235\,s, respectively, which are marked in Figure \ref{fig:search}. We also marked the spin period they reported, which is close to the ($\omega_x - \Omega$) beat period of X-rays. The harmonic of the optical period of 119\,s is not detected in the FXT data.

\begin{table*}
\caption{
\label{tab:par}
Spectral fitting parameters with C-statistic }
\centering
\begin{tabular}{l l l l l l}
\hline\hline
Component& Parameter       & Unit                   & Average                 & Pulse-on(0.2-0.6)      & Pulse-off(0.6-1.2)\\
\hline
tbabs    & $n_{\rm H}$     & $10^{22}\,\rm cm^{-2}$ & 0.086(fixed)            & 0.086(fixed)           & 0.086(fixed)      \\ 
tbpcf   & $n_{\rm H}$     & $10^{22}\,\rm cm^{-2}$ & 2.0$_{-0.4}^{+0.5}$     & 1.7$_{-0.4}^{+0.4}$    & 2.9$_{-0.5}^{+0.6}$\\
         & $f_{\rm cover}$ &                        & 0.87$_{-0.04}^{+0.02}$  & 0.88$_{-0.03}^{+0.03}$ & 0.91$_{-0.02}^{+0.02}$\\
bremss   & $kT$            & keV                    & \textgreater\,11   & 12(fixed)              & 12(fixed) \\
         & $norm$          & $10^{-3}$              & 3.3$_{-0.3}^{+0.2}$     & 3.5$_{-0.4}^{+0.5}$    & 3.6$_{-0.4}^{+0.5}$\\

Fitting  & C-stat (d.o.f.) &                        & 20(26)                  & 29(27)                 & 25(26)\\
Flux     & Absorbed        & $10^{-11}$\,ergs\,cm$^{-2}$\,s$^{-1}$ & 1.32$_{-0.35}^{+0.04}$ & 1.33$_{-0.08}^{+0.12}$ & 1.26$_{-0.08}^{+0.07}$\\
(0.3$-$10\,keV)& Unabsorbed& $10^{-11}$\,ergs\,cm$^{-2}$\,s$^{-1}$ & 1.81$_{-0.15}^{+0.18}$ & 1.93$_{-0.23}^{+0.27}$ & 1.99$_{-0.24}^{+0.27}$\\
\hline
\hline
\end{tabular}
\end{table*}

The pulse profile in the range of $0.3-2$\,keV is folded with a period of 231\,s with phase 0 corresponding to the midpoint of the observation (inset of Figure \ref{fig:search}). The folded pulse profile exhibits a single peak per cycle (Figure \ref{fig:search}), similar to that in the optical band. Without a spin ephemeris in the optical band, a direct phase comparison between the X-ray and optical pulse profiles is not possible. The pulse fraction in $0.3-2$\,keV is calculated to be $41 \pm 9\,\%$, defined as ($C_{\rm max}-C_{\rm min})/(C_{\rm max}+C_{\rm min})$, where $C_{\rm max}$ and $C_{\rm min}$ represent the maximum and minimum counts in the folded pulse profile. The pulse fraction of the 2-10 keV profile (folded with a period of 231\,s) is only $20 \pm 10\,\%$. The spin modulation amplitude usually varies with energy for IPs in the X-ray band. Most of them have stronger modulation at lower energies due to the photoelectric absorption effects from the pre-shock material within the accretion curtain \citep{1988MNRAS.231..549R, 1993MNRAS.265L..35H,2017PASP..129f2001M,2020AdSpR..66.1209D}. The absence of periodic variations in the 2–10\,keV band is consistent with this interpretation.

The FXT spectra are fitted using the \texttt{XSPEC} 12.14.0b software package \citep{arnaud96}. The events in energy ranges of 0.1$-$1.0\,keV, 1.0$-$4.1\,keV, 4.1$-$5.7\,keV, and 5.7$-$11.9\,keV are grouped in 15, 33, 7 and 4 bins using the tool \texttt{grppha}. We apply the model: $tbabs \times tbpcf \times bremss$, to fit the $0.3-10$\,keV spectrum generated from both the FXT-A and FXT-B data (left panel of Figure \ref{fig:spec}). The Tuebingen-Boulder model $tbabs$ \citep{2000ApJ...542..914W} is used to account for the interstellar medium absorption along the line of sight. The parameter $n_{\rm H}$ of $tbabs$ is fixed at the Galactic value of 8.6$\times10^{20}\,{\rm cm^{-2}}$ from \citealt{2016A&A...594A.116H}. The $tbpcf$ model, which is typically used in the spectral fitting of IPs \citep{1989MNRAS.237..853N,2017PASP..129f2001M}, is included to account for the strong absorption below 2\,keV. The inclusion of a partial covering absorber reduced the C-stat by 101 (with 26 d.o.f.) thus accounting for localised absorption of the X-ray emission by the pre-shock accretion flow \citep{1998MNRAS.298..737D}. We find that the local partial absorption has a hydrogen column density of 2.0$\times10^{22}\,{\rm cm^{-2}}$ and a covering fraction of 88\%. The X-ray radiation from post-shock region can be described by the bremsstrahlung process. Due to the limitation of the energy range of FXT, the fit yields a lower limit of 11\,keV for the plasma temperature. We applied the optically thin emitting plasma model, $apec$, for comparison. We found that it was also unable to constrain the plasma temperature, and only reduced the C-stat by 0.18 (with 26 d.o.f.). To avoid over-interpreting the model, we still use $bremss$ model for fitting. Using the distance $d = 309.5\pm0.4\,$pc of the optical counterpart, Gaia DR3 5370642890382757888, derived by \cite{2021AJ....161..147B}, the unabsorbed luminosity in the 0.3$-$10\,keV energy range is $\sim 2.1\times10^{32}$ erg\,s$^{-1}$. The fitting results are listed in Table \ref{tab:par}. Errors are quoted at the 90\% confidence level for a single parameter of interest.

The phase-resolved spectra are extracted by collecting the source events into "pulse-on" and "pulse-off" parts using the \texttt{Xselect} tool (see the inset in Figure \ref{fig:search}). The same model is applied to fit the pulse-on (phase $0.2-0.6$) spectrum and the pulse-off (phase $0.6-1.2$) spectrum (Figure \ref{fig:spec}, right panel). We fix the plasma temperature of the bremsstrahlung model component at 12\,keV due to the low statistic at higher energies. We have checked that the constraints of other fitting parameters are robust against changes in the plasma temperature. The main spectral differences between the pulse-on and pulse-off intervals occur below 3\,keV. In terms of fitting parameters, this is reflected in the variation of the column density of the partial covering absorption component (Table \ref{tab:par}) as expected in IP systems. The normalization of the bremsstrahlung component displays no significant variation within error.
 
%%%%%%%%%%%%%%%%%%%%%%%%%%%%%%%%%%%%%%%%%%

\section{Radio upper limits}
\label{sec:Radio}

The Australian SKA Pathfinder \citep[ASKAP;][]{2021PASA...38....9H} is a 36-antenna interferometer equipped with phased-array feeds that widen its field-of-view to approximately 36~sq.~deg, observing at a frequency range of 800--1500\,MHz. It is undertaking several large surveys, the calibrated visibilities of which are accessible shortly after observation\footnote{via the CSIRO Astronomy Science Data Access portal: \url{https://data.csiro.au/domain/casdaObservation/}}. \src{} was within the field-of-view of observations from the Variables and Slow Transients \citep[VAST;][]{Murphy2013} survey and the Rapid ASKAP Continuum Survey \citep[RACS;][]{2020PASA...37...48M}, which have integration times of 12\,min and 15\,min per observation, respectively. We searched for five epochs of RACS spaced over 2020-03-27 to 2024-01-11, and nine epochs of VAST spaced over 2023-06-23 to 2024-08-16. We examined images formed from the integration of each epoch \citep[via \textsc{WSClean;}][]{2014MNRAS.444..606O}, finding no counterparts down to RMS noise levels of 200--300\,$\mu$Jy\,beam$^{-1}$. We also performed a time-domain search at a resolution of 10\,s, by subtracting the continuum model and re-imaging every time step. We found no bursting transient or pulsed radio signals down to RMS noise levels of $\sim2$--3\,mJy\,beam$^{-1}$.

The Murchison Widefield Array \citep[MWA;][]{2013PASA...30....7T,2018PASA...35...33W} is a low-frequency (80--300\,MHz), wide-field-of-view (500--1,000\,sq.\,deg.) radio interferometer comprising 128--256\,tiles, co-located with ASKAP. Under project code G0080, the Galactic Plane Monitor has been repeatedly scanning the Southern Milky Way over $|b|<15^\circ$ (with decreasing sensitivity from $|b|>10^\circ$) with integrations of 30--45\,minutes at 185--215\,MHz from 2022 to 2025 \citep[see Methods of][; to be described in full by Hurley-Walker et al. in prep]{2023Natur.619..487H}. 11~observations with sensitivity to \src{} were taken from 2024-12-15 to 2025-01-06. No counterparts were found in continuum imaging at five-minute intervals down to a noise level of 40--90\,mJy\,beam$^{-1}$. A time-domain search following the same method as for the ASKAP data yielded no bursting transient or pulsed radio signals on a 4-s cadence to RMS noise levels of $\sim$350--800\,mJy\,beam$^{-1}$.

\section{Discussion and conclusion}
\label{sec:conclusion}

\src\ is an X-ray transient source discovered by the EP mission. The optical follow-up observations \citep{2024MNRAS.532L..21P} classified this source as an IP. Optical photometry and spectroscopy revealed a binary orbital period of 3.7614\,hr and a proposed WD spin period of 238\,s, while the optical beat period of 243\,s was also detected. Using the FXT observation, a periodic modulation at a period of 231$\pm2$\,s was detected in the energy range of 0.3$-$2\,keV, with a significance exceeding 4\,$\sigma$. However, no harmonics of the spin period were detected as shown in Figure \ref{fig:search}. The broad envelope of the peak in the $Z_{2}^{2}$ periodogram precludes a clear determination of a beat period in the X-rays.

Our spin periodicity detection in the X-ray band deviates significantly from those reported in the optical band, and no periodic signal corresponding to 231\,s was found in the Lomb–Scargle Periodogram in \citealt{2024MNRAS.532L..21P}. However, the closeness of the longer optical period of 238\,s to the possible negative beat period ($\omega_x - \Omega$) of 235\,s in X-rays, considering the relatively large uncertainty of the X-ray periodicity estimated at 2\,s at the 68$\%$ level, may help reconcile the optical and X-ray results. If the peak at 362.5\,c/d observed in the optical band corresponds to the beat frequency, using the orbital frequency derived from TESS of 6.3806\,c/d, then the spin frequency is 368.9\,c/d corresponding to a spin period of 234\,s, which is within 2$\sigma$ of the inferred X-ray period. The additional peak observed in the optical periodogram and interpreted as the beat frequency would instead correspond to ($\omega - 2\Omega$) at 356.12\,c/d or 242.6\,s. The differing relative dominance of the spin and beat frequencies between the X-ray and optical bands is not unusual in IPs (see e.g. \citep{2012A&A...542A..22B}. This can be explained by the fact that optical emission is affected by X-ray reprocessing at different locations within the binary system.

Our spectral analysis reveals that the X-ray emission from this source is thermal and well modeled by an bremsstrahlung model whose temperature ($>11$\,keV) cannot be constrained with the present data and affected by strong partial ($\sim 90\%$) covering absorption below 2\,keV. The local hydrogen absorption column density reaches $2.0\times 10^{22}\,\rm cm^{-2}$, significantly exceeding the interstellar value of $8.6\times 10^{20}\,\rm cm^{-2}$ estimated based on \citealt{2016A&A...594A.116H}. The phase-resolved spectral analysis shows that the periodic variation detected in the energy range of 0.3$-$2\,keV is mainly caused by the change in the absorption column density. This may indicate that the cool pre-shock accretion flow absorbs the X-ray emission at different angles as the WD rotates. 

The EP/WXT long term coverage shows \src\, undergoing luminosity changes of at least a factor of three over about a month timespan. This is also confirmed by the long term optical photometric coverage reported by \cite{2024MNRAS.532L..21P}. IPs rarely undergo frequent luminosity state changes (e.g. \citep{2017MNRAS.469..956K,2025ATel17050....1L}, indicating an unusual behaviour of this newly discovered magnetic system.

In the past few years, a new class of objects have been discovered emitting bright periodic radio bursts, called Long Period Transients (LPTs). Eight sources have been discovered by now, but the recent discovery of two LPTs in M dwarf-WD binary systems (ILT~J1101+5521 and GLEAM-X~J0704--37) suggests that WD binaries, especially with short orbital periods, might be progenitors for some of the LPTs. In that view we searched in the radio archive for \src\, but could only derive an upper limit. 
Thus despite its unusual behaviour \src\, does not appear to belong to this new class of systems.

The detection of transient X-ray emission from \src\ with EP, along with subsequent X-ray follow-up and coordinated optical observations, has enabled us to classify this system as an IP. This demonstrates the potential of the EP mission to discover previously unknown CV systems that re-emerge from deep low states or undergo outbursts, particularly IPs whose state changes are extremely rare.

\begin{acknowledgements}
 This work is based on data obtained with Einstein Probe, a space mission supported by Strategic Priority Program on Space Science of Chinese Academy of Sciences, in collaboration with ESA, MPE and CNES (Grant No. XDA15310000), the Strategic Priority Research Program of the Chinese Academy of Sciences (Grant No. XDB0550200), and the National Key R\&D Program of China (2022YFF0711500). 
We acknowledge the support by the National Natural Science Foundation of China (Grant Nos. 12321003, 12103065, 12333004, 12373040, 12021003, 12473016, 12373051, 12273029, 12221003), the China Manned Space Project (Grant Nos. CMS-CSST-2021-A13, CMS-CSST-2021-B11), and the Youth Innovation Promotion Association of the Chinese Academy of Sciences. We acknowledge the data resources and technical support provided by the China National Astronomical Data Center, the Astronomical Science Data Center of the Chinese Academy of Sciences, and the Chinese Virtual Observatory. This scientific work uses data obtained from Inyarrimanha Ilgari Bundara, the CSIRO Murchison Radio-astronomy Observatory. Support for the operation of the MWA is provided by the Australian Government (NCRIS), under a contract to Curtin University administered by Astronomy Australia Limited.
The Australian SKA Pathfinder is part of the Australia Telescope National Facility which is managed by CSIRO. Operation of ASKAP is funded by the Australian Government with support from the National Collaborative Research Infrastructure Strategy. ASKAP and the MWA use the resources of the Pawsey Supercomputing Centre. Establishment of ASKAP, Inyarrimanha Ilgari Bundara, and the Pawsey Supercomputing Centre are initiatives of the Australian Government, with support from the Government of Western Australia and the Science and Industry Endowment Fund. We acknowledge the Wajarri Yamaji People as the Traditional Owners and Native Title Holders of the observatory site. N.R. is supported by the European Research Council (ERC) via the Consolidator Grant “MAGNESIA” (No. 817661) and the Proof of Concept ``DeepSpacePulse" (No. 101189496), by the Catalan grant SGR2021-01269, the Spanish grant ID2023-153099NA-I00, and by the program Unidad de Excelencia Maria de Maeztu CEX2020-001058-M. 
DdM acknowledges financial support from INAF Research Grants AstroFund 2022 and 2024. FCZ is supported by a Ram\'on y Cajal fellowship (grant agreement RYC2021-030888-I). NHW is supported by an Australian Research Council Future Fellowship (project number FT190100231) funded by the Australian Government. Support for KM is provided by NASA ADAP program (NNH22ZDA001N-ADAP) grant.

\end{acknowledgements}

\bibliographystyle{aa}
\bibliography{aa54876-25}

\begin{thebibliography}{29}
\expandafter\ifx\csname natexlab\endcsname\relax\def\natexlab#1{#1}\fi

\bibitem[{{Arnaud}(1996)}]{arnaud96}
{Arnaud}, K.~A. 1996, in Astronomical Society of the Pacific Conference Series, Vol. 101, Astronomical Data Analysis Software and Systems V, ed. G.~H. {Jacoby} \& J.~{Barnes}, 17

\bibitem[{{Bailer-Jones} {et~al.}(2021){Bailer-Jones}, {Rybizki}, {Fouesneau}, {Demleitner}, \& {Andrae}}]{2021AJ....161..147B}
{Bailer-Jones}, C.~A.~L., {Rybizki}, J., {Fouesneau}, M., {Demleitner}, M., \& {Andrae}, R. 2021, \aj, 161, 147

\bibitem[{{Bernardini} {et~al.}(2012){Bernardini}, {de Martino}, {Falanga}, {Mukai}, {Matt}, {Bonnet-Bidaud}, {Masetti}, \& {Mouchet}}]{2012A&A...542A..22B}
{Bernardini}, F., {de Martino}, D., {Falanga}, M., {et~al.} 2012, \aap, 542, A22

\bibitem[{{Bernardini} {et~al.}(2018){Bernardini}, {de Martino}, {Mukai}, \& {Falanga}}]{2018MNRAS.478.1185B}
{Bernardini}, F., {de Martino}, D., {Mukai}, K., \& {Falanga}, M. 2018, \mnras, 478, 1185

\bibitem[{{Buccheri} {et~al.}(1983){Buccheri}, {Bennett}, {Bignami}, {Bloemen}, {Boriakoff}, {Caraveo}, {Hermsen}, {Kanbach}, {Manchester}, {Masnou}, {Mayer-Hasselwander}, {{\"O}zel}, {Paul}, {Sacco}, {Scarsi}, \& {Strong}}]{1983A&A...128..245B}
{Buccheri}, R., {Bennett}, K., {Bignami}, G.~F., {et~al.} 1983, \aap, 128, 245

\bibitem[{{Buckley} {et~al.}(2024){Buckley}, {Monageng}, {Aydi}, {Scaringi}, \& {Charles}}]{2024ATel16554....1B}
{Buckley}, D.~A.~H., {Monageng}, I., {Aydi}, E., {Scaringi}, S., \& {Charles}, P.~A. 2024, The Astronomer's Telegram, 16554, 1

\bibitem[{{Chang} {et~al.}(2024){Chang}, {Jiang}, {Liao}, {Cui}, {Huang}, {An}, {Cao}, \& {Xu}}]{2024ATel16572....1C}
{Chang}, N., {Jiang}, P., {Liao}, J., {et~al.} 2024, The Astronomer's Telegram, 16572, 1

\bibitem[{{de Martino} {et~al.}(2020){de Martino}, {Bernardini}, {Mukai}, {Falanga}, \& {Masetti}}]{2020AdSpR..66.1209D}
{de Martino}, D., {Bernardini}, F., {Mukai}, K., {Falanga}, M., \& {Masetti}, N. 2020, Advances in Space Research, 66, 1209

\bibitem[{{Done} \& {Magdziarz}(1998)}]{1998MNRAS.298..737D}
{Done}, C. \& {Magdziarz}, P. 1998, \mnras, 298, 737

\bibitem[{{Hellier}(1993)}]{1993MNRAS.265L..35H}
{Hellier}, C. 1993, \mnras, 265, L35

\bibitem[{{HI4PI Collaboration} {et~al.}(2016){HI4PI Collaboration}, {Ben Bekhti}, {Fl{\"o}er}, {Keller}, {Kerp}, {Lenz}, {Winkel}, {Bailin}, {Calabretta}, {Dedes}, {Ford}, {Gibson}, {Haud}, {Janowiecki}, {Kalberla}, {Lockman}, {McClure-Griffiths}, {Murphy}, {Nakanishi}, {Pisano}, \& {Staveley-Smith}}]{2016A&A...594A.116H}
{HI4PI Collaboration}, {Ben Bekhti}, N., {Fl{\"o}er}, L., {et~al.} 2016, \aap, 594, A116

\bibitem[{{Hotan} {et~al.}(2021){Hotan}, {Bunton}, {Chippendale}, {Whiting}, {Tuthill}, {Moss}, {McConnell}, {Amy}, {Huynh}, {Allison}, {Anderson}, {Bannister}, {Bastholm}, {Beresford}, {Bock}, {Bolton}, {Chapman}, {Chow}, {Collier}, {Cooray}, {Cornwell}, {Diamond}, {Edwards}, {Feain}, {Franzen}, {George}, {Gupta}, {Hampson}, {Harvey-Smith}, {Hayman}, {Heywood}, {Jacka}, {Jackson}, {Jackson}, {Jeganathan}, {Johnston}, {Kesteven}, {Kleiner}, {Koribalski}, {Lee-Waddell}, {Lenc}, {Lensson}, {Mackay}, {Mahony}, {McClure-Griffiths}, {McConigley}, {Mirtschin}, {Ng}, {Norris}, {Pearce}, {Phillips}, {Pilawa}, {Raja}, {Reynolds}, {Roberts}, {Roxby}, {Sadler}, {Shields}, {Schinckel}, {Serra}, {Shaw}, {Sweetnam}, {Troup}, {Tzioumis}, {Voronkov}, \& {Westmeier}}]{2021PASA...38....9H}
{Hotan}, A.~W., {Bunton}, J.~D., {Chippendale}, A.~P., {et~al.} 2021, \pasa, 38, e009

\bibitem[{{Hurley-Walker} {et~al.}(2023){Hurley-Walker}, {Rea}, {McSweeney}, {Meyers}, {Lenc}, {Heywood}, {Hyman}, {Men}, {Clarke}, {Coti Zelati}, {Price}, {Horv{\'a}th}, {Galvin}, {Anderson}, {Bahramian}, {Barr}, {Bhat}, {Caleb}, {Dall'Ora}, {de Martino}, {Giacintucci}, {Morgan}, {Rajwade}, {Stappers}, \& {Williams}}]{2023Natur.619..487H}
{Hurley-Walker}, N., {Rea}, N., {McSweeney}, S.~J., {et~al.} 2023, \nat, 619, 487

\bibitem[{{Kennedy} {et~al.}(2017){Kennedy}, {Garnavich}, {Littlefield}, {Callanan}, {Mukai}, {Aadland}, {Kotze}, \& {Kotze}}]{2017MNRAS.469..956K}
{Kennedy}, M.~R., {Garnavich}, P.~M., {Littlefield}, C., {et~al.} 2017, \mnras, 469, 956

\bibitem[{{Ling} {et~al.}(2024){Ling}, {Liu}, {Liu}, {Jin}, {Zhang}, {Cheng}, {Chen}, {Cui}, {Fan}, {Hu}, {Hu}, {Huang}, {Li}, {Lian}, {Liu}, {Lv}, {Mao}, {Pan}, {Pan}, {Sun}, {Wang}, {Wang}, {Wu}, {Xu}, {Xu}, {Yang}, {Zhang}, {Zhang}, {Zhang}, {Zhang}, {Zhao}, {Chen}, {Cui}, {Feng}, {Guan}, {Han}, {Jia}, {Li}, {Li}, {Liu}, {Lu}, {Song}, {Wang}, {Xu}, {Zhang}, {Zhang}, {Zhao}, {Zhao}, {Kuulkers}, {Santovincenzo}, {Saxton}, {O'Brien}, {Rau}, {Nandra}, {Friedrich}, {Meidinger}, {Burwitz}, {Cordier}, {Rea}, \& {Yuan}}]{2024ATel16546....1L}
{Ling}, Z.~X., {Liu}, M.~J., {Liu}, Y., {et~al.} 2024, The Astronomer's Telegram, 16546, 1

\bibitem[{{Littlefield} {et~al.}(2025){Littlefield}, {Bonnardeau}, \& {Garnavich}}]{2025ATel17050....1L}
{Littlefield}, C., {Bonnardeau}, M., \& {Garnavich}, P. 2025, The Astronomer's Telegram, 17050, 1

\bibitem[{{McConnell} {et~al.}(2020){McConnell}, {Hale}, {Lenc}, {Banfield}, {Heald}, {Hotan}, {Leung}, {Moss}, {Murphy}, {O'Brien}, {Pritchard}, {Raja}, {Sadler}, {Stewart}, {Thomson}, {Whiting}, {Allison}, {Amy}, {Anderson}, {Ball}, {Bannister}, {Bell}, {Bock}, {Bolton}, {Bunton}, {Chippendale}, {Collier}, {Cooray}, {Cornwell}, {Diamond}, {Edwards}, {Gupta}, {Hayman}, {Heywood}, {Jackson}, {Koribalski}, {Lee-Waddell}, {McClure-Griffiths}, {Ng}, {Norris}, {Phillips}, {Reynolds}, {Roxby}, {Schinckel}, {Shields}, {Tremblay}, {Tzioumis}, {Voronkov}, \& {Westmeier}}]{2020PASA...37...48M}
{McConnell}, D., {Hale}, C.~L., {Lenc}, E., {et~al.} 2020, \pasa, 37, e048

\bibitem[{{Mukai}(2017)}]{2017PASP..129f2001M}
{Mukai}, K. 2017, \pasp, 129, 062001

\bibitem[{{Murphy} {et~al.}(2013){Murphy}, {Chatterjee}, {Kaplan}, {Banyer}, {Bell}, {Bignall}, {Bower}, {Cameron}, {Coward}, {Cordes}, {Croft}, {Curran}, {Djorgovski}, {Farrell}, {Frail}, {Gaensler}, {Galloway}, {Gendre}, {Green}, {Hancock}, {Johnston}, {Kamble}, {Law}, {Lazio}, {Lo}, {Macquart}, {Rea}, {Rebbapragada}, {Reynolds}, {Ryder}, {Schmidt}, {Soria}, {Stairs}, {Tingay}, {Torkelsson}, {Wagstaff}, {Walker}, {Wayth}, \& {Williams}}]{Murphy2013}
{Murphy}, T., {Chatterjee}, S., {Kaplan}, D.~L., {et~al.} 2013, \pasa, 30, e006

\bibitem[{{Norton} \& {Watson}(1989)}]{1989MNRAS.237..853N}
{Norton}, A.~J. \& {Watson}, M.~G. 1989, \mnras, 237, 853

\bibitem[{{Offringa} {et~al.}(2014){Offringa}, {McKinley}, {Hurley-Walker}, {Briggs}, {Wayth}, {Kaplan}, {Bell}, {Feng}, {Neben}, {Hughes}, {Rhee}, {Murphy}, {Bhat}, {Bernardi}, {Bowman}, {Cappallo}, {Corey}, {Deshpande}, {Emrich}, {Ewall-Wice}, {Gaensler}, {Goeke}, {Greenhill}, {Hazelton}, {Hindson}, {Johnston-Hollitt}, {Jacobs}, {Kasper}, {Kratzenberg}, {Lenc}, {Lonsdale}, {Lynch}, {McWhirter}, {Mitchell}, {Morales}, {Morgan}, {Kudryavtseva}, {Oberoi}, {Ord}, {Pindor}, {Procopio}, {Prabu}, {Riding}, {Roshi}, {Shankar}, {Srivani}, {Subrahmanyan}, {Tingay}, {Waterson}, {Webster}, {Whitney}, {Williams}, \& {Williams}}]{2014MNRAS.444..606O}
{Offringa}, A.~R., {McKinley}, B., {Hurley-Walker}, N., {et~al.} 2014, \mnras, 444, 606

\bibitem[{{Parker} {et~al.}(2005){Parker}, {Norton}, \& {Mukai}}]{2005A&A...439..213P}
{Parker}, T.~L., {Norton}, A.~J., \& {Mukai}, K. 2005, \aap, 439, 213

\bibitem[{{Patterson}(1994)}]{1994PASP..106..209P}
{Patterson}, J. 1994, \pasp, 106, 209

\bibitem[{{Potter} {et~al.}(2024){Potter}, {Buckley}, {Scaringi}, {Monageng}, {Egbo}, {Charles}, {Erasmus}, {van Gend}, {Loubser}, {Titus}, {Rosie}, {Gajjar}, {Worters}, {Chandra}, {Julie}, \& {Hlakola}}]{2024MNRAS.532L..21P}
{Potter}, S.~B., {Buckley}, D. A.~H., {Scaringi}, S., {et~al.} 2024, \mnras, 532, L21

\bibitem[{{Rosen} {et~al.}(1988){Rosen}, {Mason}, \& {Cordova}}]{1988MNRAS.231..549R}
{Rosen}, S.~R., {Mason}, K.~O., \& {Cordova}, F.~A. 1988, \mnras, 231, 549

\bibitem[{{Tingay} {et~al.}(2013){Tingay}, {Goeke}, {Bowman}, {Emrich}, {Ord}, {Mitchell}, {Morales}, {Booler}, {Crosse}, {Wayth}, {Lonsdale}, {Tremblay}, {Pallot}, {Colegate}, {Wicenec}, {Kudryavtseva}, {Arcus}, {Barnes}, {Bernardi}, {Briggs}, {Burns}, {Bunton}, {Cappallo}, {Corey}, {Deshpande}, {Desouza}, {Gaensler}, {Greenhill}, {Hall}, {Hazelton}, {Herne}, {Hewitt}, {Johnston-Hollitt}, {Kaplan}, {Kasper}, {Kincaid}, {Koenig}, {Kratzenberg}, {Lynch}, {Mckinley}, {Mcwhirter}, {Morgan}, {Oberoi}, {Pathikulangara}, {Prabu}, {Remillard}, {Rogers}, {Roshi}, {Salah}, {Sault}, {Udaya-Shankar}, {Schlagenhaufer}, {Srivani}, {Stevens}, {Subrahmanyan}, {Waterson}, {Webster}, {Whitney}, {Williams}, {Williams}, \& {Wyithe}}]{2013PASA...30....7T}
{Tingay}, S.~J., {Goeke}, R., {Bowman}, J.~D., {et~al.} 2013, 30, 7

\bibitem[{{Wayth} {et~al.}(2018){Wayth}, {Tingay}, {Trott}, {Emrich}, {Johnston-Hollitt}, {McKinley}, {Gaensler}, {Beardsley}, {Booler}, {Crosse}, {Franzen}, {Horsley}, {Kaplan}, {Kenney}, {Morales}, {Pallot}, {Sleap}, {Steele}, {Walker}, {Williams}, {Wu}, {Cairns}, {Filipovic}, {Johnston}, {Murphy}, {Quinn}, {Staveley-Smith}, {Webster}, \& {Wyithe}}]{2018PASA...35...33W}
{Wayth}, R.~B., {Tingay}, S.~J., {Trott}, C.~M., {et~al.} 2018, 35, 33

\bibitem[{{Wilms} {et~al.}(2000){Wilms}, {Allen}, \& {McCray}}]{2000ApJ...542..914W}
{Wilms}, J., {Allen}, A., \& {McCray}, R. 2000, \apj, 542, 914

\bibitem[{{Yuan} {et~al.}(2022){Yuan}, {Zhang}, {Chen}, \& {Ling}}]{2022hxga.book...86Y}
{Yuan}, W., {Zhang}, C., {Chen}, Y., \& {Ling}, Z. 2022, in Handbook of X-ray and Gamma-ray Astrophysics, ed. C.~{Bambi} \& A.~{Sangangelo}, 86

\end{thebibliography}

\end{document}